\documentclass[12pt]{article}

 \usepackage{amssymb}
 \usepackage{latexsym}

\def\barr{\left(\begin{array}}
\def\earr{\end{array}\right)}

\newcommand{\beq}[1]{\begin{equation}\label{#1}}
\newcommand{\bear}[1]{\begin{eqnarray}\label{#1}}

\newcommand{\eeq}{\end{equation}}
\newcommand{\ear}{\end{eqnarray}}

 \renewcommand{\theequation}{\arabic{section}.\arabic{equation}}
 \catcode`\@=11 \@addtoreset{equation}{section}\catcode`\@=12

\newcommand{\R}{ {\mathbb R} }

\newcommand{\btu}{\bigtriangleup}

\newcommand{\p}{\partial}

\newcommand{\fnm}{\footnotemark}
\newcommand{\fnt}{\footnotetext}

\newcommand{\nn}{ {\nonumber } }

\begin{document}

 \begin{center}
 \large\bf Electric S-brane solutions corresponding to rank-2  Lie
  algebras:  acceleration and small variation of G
 \\[12pt]
    V.D. Ivashchuk\fnm[1]\fnt[1]{ivashchuk@mail.ru}$^{, \ a, \ b}$,
    S.A. Kononogov\fnm[2]\fnt[2]{kononogov@vniims.ru}$^{, \ a}$ \\
     and V.N. Melnikov\fnm[3]\fnt[3]{melnikov@phys.msu.ru}$^{, \ a, \ b}$\\

 \vspace{3pt}

 \it
 $a \quad$ Center for Gravitation and Fundamental Metrology,
  VNIIMS, 46 Ozyornaya St., Moscow 119361, Russia  \\

 $b \quad$ Institute of Gravitation and Cosmology,
  Peoples' Friendship University of Russia,
  6 Miklukho-Maklaya St., \\
  Moscow 117198, Russia

 \end{center}

 \vspace{15pt}

 \begin{abstract}

 Electric $S$-brane solutions with two non-composite electric
 branes and a set of $l$ scalar fields are considered. The
 intersection rules for branes correspond to Lie algebras
 $A_2$, $C_2$ and $G_2$. The solutions contain five factor spaces.
 One of them, $M_0$, is interpreted as our 3-dimensional space.
 It is shown that there exists a time interval where
 accelerated expansion of our 3-dimensional space is
 compatible with a small enough variation of the effective gravitational
 constant $G(\tau)$. This interval contains
 $\tau_0$,  a point of minimum of the function $G(\tau)$.
 A special  solution with two phantom scalar fields is analyzed and
  it is shown that in the vicinity of the point  $\tau_0$
  the time variation of $G(\tau)$ (calculated in the linear approximation)
  decreases  in the sequence of Lie algebras $A_{2}$, $C_{2}$ and $G_2$.

\end{abstract}

\pagebreak

\section{Introduction}

 Multidimensional cosmological models with diverse
matter sources are at present widely used  for describing possible
time variations of fundamental physical constants, e.g.
gravitational constant $G$, see  \cite{Mel2}-\cite{AIM-07} and
references therein.

 It has been shown in \cite{DIKM} that, in the pure gravitational model
 with two non-zero curvatures,  there exists an interval of
 synchronous variable $\tau$  where accelerated expansion of
 ``our'' 3-dimensional space co-exists with a small enough value of
 $\dot G$. This result was compared with our exact
 (1+3+6)-dimensional solution \cite{GIM} obtained earlier.

Recently, in \cite{AIM-07}, we suggested  a similar mechanism for
a model with two form fields and several scalar fields (e.g.,
phantom ones). The main problem here was to find  an interval of
 the synchronous time $\tau$ where the scale factor
of our 3-dimensional space exhibits an accelerated expansion
according to the observational data \cite{Riess,Perl} while the
relative variation of the effective 4-dimensional gravitational
constant is small enough as compared with the Hubble parameter,
see \cite{Hel,Dic,BZhuk} and references therein. As it was shown
in \cite{AIM-07} such interval did exist,  it contained
 $\tau_0$ -- the point of minimum of the function $G(\tau)$.
The analysis carried out in  \cite{AIM-07} was based on an exact
 $S$-brane solution with the intersection rules for branes corresponding
to the Lie algebra  $A_2$.

 In this paper, we extend the results of our previous work to exact
 $S$-brane solutions with the intersection rules corresponding
 to other simple rank 2 Lie algebras:  $C_2$ and $G_2$. Thus,
 here we generalize the results of ref. \cite{AIM-07} to the sequence of
 Lie algebras $A_{2}$, $C_{2}$ and $G_2$.

 The paper is organized as follows. In Section 2, the
 setup for the model is done and exact $S$-brane solutions
 are presented.
 In Section 3,  solutions with acceleration and small G-dot
 are singled out. In Section 4, a special configuration with two
 phantom field is considered.  Here we compare the
 $G$-dot  calculated (in the linear approximation near the point
 of minimum of $G(\tau)$)  for  these three algebras
 and show that the variation of $G$ decreases
 in the sequence of Lie algebras  $A_{2}$, $C_{2}$ and $G_2$.
 In the Appendix,  we give a derivation of  relation
 (\ref{2.13}) for an approximate value of the dimensionless
 parameter of  relative $G$ variation.

\section{The model}

 We consider  $S$-brane solutions describing two electric branes
 and a set of $l$ scalar fields.

The model is governed by the action
 \begin{eqnarray}
   S=\int d^Dx \sqrt{|g|} \biggl\{
    R[g]- h_{\alpha\beta} g^{MN}\p_M\varphi^\alpha \p_N\varphi^\beta
    \nn    \\ \label{1.1}
  -\sum_{a = 1,2}\frac{1}{N_a!} \exp[2\lambda_a(\varphi)](F^a)^2 \biggr\},
 \end{eqnarray}
where  $g=g_{MN}(x)dx^M\otimes dx^N$ is a $D$-dimensional metric
of the pseudo-Euclidean signature $(-,+, \dots, +)$, $F^a = dA^a$
is a form of rank $N_a$, $(h_{\alpha\beta})$ is a non-degenerate
symmetric matrix, $\varphi=(\varphi^\alpha) \in
 \R^l$ is a vector of $l$ scalar fields,
 $\lambda_a(\varphi)=\lambda_{a \alpha}\varphi^\alpha$ is a linear
function, with $a = 1,2$ and $\alpha, \beta =1, \dots, l$ and $|g|
 = |\det (g_{MN})|$.

We consider the manifold
 \beq{1.2}
    M =    (0, + \infty)  \times M_{0} \times M_1 \times M_2 \times
        M_3 \times M_4 .
\eeq
where $M_i $ are oriented Riemannian Ricci-flat spaces of
dimensions $d_i$, $i = 0,  \dots, 4$, and $d_1 = 1$.

Let two electric branes be defined by the sets $I_1 = \{ 1, 2, 3 \}$ and
$I_2 = \{ 1, 2, 4 \}$. They intersect on $M_1 \times M_2$. The first brane
also covers $M_3$ while the second one covers $M_4$. The first brane
corresponds to the form $F^1$ and the second one to the form $F^2$.

For the world-volume dimensions of branes we get
 \beq{1.3}
    d(I_s) = N_s -1 = 1 + d_2 + d_{2 + s},
 \eeq
 $s=1,2$, and
  \beq{1.4}
     d(I_1 \cap I_2) = 1 + d_2
  \eeq
is the brane intersection dimension.

We consider an $S$-brane solution governed by the function
 \beq{1.5}
      \hat{H} = 1 + P t^2,
 \eeq
where $t$ is a time variable and
 \beq{1.6}
    P = \frac{1}{4 n_s} K_s Q_s^2
\eeq
 is a parameter. Here, $Q_s$ are charge density parameters,
  \beq{1.7}
  K_{s} =   d(I_s)\bigl( 1+ \frac{d(I_s)}{2-D}\bigr)
      + \lambda_{s \alpha }\lambda_{s \beta} h^{\alpha \beta},
 \eeq
  $s = 1,2$,  $(h^{\alpha \beta}) = (h_{\alpha \beta})^{-1}$,
 and
    \beq{1.7n}
         (n_1, n_2) = (2,2), (3,4), (6,10)
       \eeq
 for the Lie algebras  $A_{2}$, $C_{2}$ and $G_2$, respectively.
 The parameters $K_s$ and $Q_s$ are supposed to be nonzero.

The intersection rules read:
  \beq{1.10}
     d(I_1 \cap I_{2}) = \frac{d(I_1)d(I_{2})}{D -2}  -
     \lambda_{1 \alpha }\lambda_{2 \beta } h^{\alpha\beta}- \frac{1}{2} K_2.
  \eeq

These relations corresponds to the simple Lie algebras  of rank 2
 \cite{IMJ,IMtop}.

 Recall that $K_s = (U^s,U^s)$, $s = 1,2$, where
the ``electric'' $U^s$-vectors and the scalar products were
defined in  \cite{IM11,IMC,IMJ}. Relations (\ref{1.10}) follow
just from the formula
 \beq{1.10a}
   (A_{s s'}) = (2(U^s,U^{s'})/(U^{s'},U^{s'})),
 \eeq
 where $(A_{s s'})$ is the Cartan matrix for  Lie algebra
 of rank 2 with  $A_{12} = -1$, $A_{21} = -k$, where
 here and in what follows
       \beq{1.9}
          k = 1, 2, 3,
       \eeq
for the  Lie algebras  $A_{2}$, $C_{2}$ and $G_2$, respectively.
 We remind the reader that (see \cite{IMC})
  \beq{1.10u}
    (U^1,U^{2}) = d(I_1 \cap I_{2}) - \frac{d(I_1)d(I_{2})}{D -2}
    +  \lambda_{1 \alpha }\lambda_{2 \beta } h^{\alpha\beta}.
 \eeq
    Due to (\ref{1.10a}) and  relations $A_{12} = -1$, $A_{21} =
    -k$, we get
  \beq{1.8}
         K_2 = K_1 k.
       \eeq

 We consider the following exact solutions
 (containing three subcases corresponding to the Lie algebras
 $A_{2}$, $C_{2}$ and $G_2$ )
 \begin{eqnarray}
    g = \hat{H}^{2 A} \biggl\{ - dt \otimes d t +  g^0
        + \hat{H}^{-2 (B_1 + B_2)} ( t^2 g^1 + g^2)
  \nn \\ \label{1.11}
   + \hat{H}^{-2 B_1}  g^3  + \hat{H}^{-2 B_2} g^4   \biggr\},
 \\   \label{1.12}
    \exp(\varphi^\alpha) = \hat{H}^{ B_1 \lambda_{1}^{\alpha}
        + B_2 \lambda_{2}^{\alpha}},
  \\ \label{1.13a}
    F^1 = - Q_1 \hat{H}^{-2n_1 + n_2}
        t dt  \wedge \tau_1 \wedge \tau_2 \wedge \tau_3, \quad
 \\ \label{1.13b}
    F^2 = - Q_2 \hat{H}^{-2n_2 + k n_1}
        t dt  \wedge \tau_1 \wedge \tau_2 \wedge \tau_4, \quad
 \end{eqnarray}
where
 \bear{1.14A}
      A =  \sum_{s = 1,2} \frac{n_s K^{-1}_s d(I_s)}{D-2},
 \\ \label{1.14B}
      B_s = n_s K^{-1}_s,
 \ear
 $s = 1,2$. Here $\tau_i$ denotes a volume form on $M_i$ ($g_1 = dx \otimes
 dx$, $\tau_1 = dx$).

  These solutions are special case of  more general solutions from
  \cite{GIM-flux} corresponding to the Lie algebras $A_2$, $C_2$ and $G_2$.
  They can also be  obtained as a special 1-block case of $S$-brane solutions from
  \cite{Is-brane}. The $A_2$-case ($k=1$) was considered in
  \cite{AIM-07}.

   We also note  that the charge density parameters $Q_s$
   obey the following relation (see (\ref{1.6})  and (\ref{1.8}))
   \beq{1.15}
    \frac{Q_1^2}{Q_2^2} = \frac{n_1}{n_2}k = 1, \ \frac{3}{2},  \ \frac{9}{5}
   \eeq
   for  the Lie algebras  $A_{2}$, $C_{2}$ and $G_2$, respectively.

 \section{Solutions with acceleration and small $\dot{G}$ }

 Let us introduce the synchronous time variable $\tau = \tau(t)$ by
 the relation:
 \beq{2.1}
  \tau = \int_{0}^{t} d \bar{t} [\hat{H}(\bar{t})]^{A}
 \eeq

We put $P < 0$, and hence due to (\ref{1.6}) all $K_s < 0$ which
implies $A < 0$. Consider two intervals of the parameter $A$:

 \bear{2.2i}
    {\bf (i)} \ \ \ A  < -1,
 \\  \label{2.2ii}
    {\bf (ii)}  \ \ -1 < A  < 0.
 \ear

In case (i), the function $\tau = \tau (t)$  monotonically
increases from $0$ to $+ \infty$, for $t \in (0, t_1)$, where $t_1
 = |P|^{-1/2}$, while in case $(ii)$ it is monotonically increases
from 0 to a finite value $\tau_1 = \tau(t_1)$.

Let the space $M_0$ be our 3-dimensional space with the scale
factor
 \beq{2.3}
      a_0 = \hat{H}^A.
 \eeq
For the first branch (i), we get the asymptotic relation
 \beq{2.4i}
    a_0 \sim  \tau^{\nu},
 \eeq
for $\tau \to +\infty $, where
\beq{2.4n}
       \nu = A/(A+1)
\eeq
and, due to (\ref{2.2i}), $\nu > 1$. For the second branch (ii) we obtain
\beq{2.4ii}
      a_0 \sim {\rm const }   (\tau_1 - \tau)^{\nu},
\eeq
for $\tau \to \tau_1 - 0$, where $\nu  < 0$ due to (\ref{2.2ii}), see
(\ref{2.4n}).

Thus, we get an asymptotic accelerated expansion of the
3-dimensional factor space $M_0$ in both cases (i) and  (ii), and
$a_0 \to + \infty$.

Moreover, it may be readily verified that the accelerated expansion takes
place for all $\tau > 0$, i.e.,
\beq{2.6}
     \dot{a}_0 > 0, \qquad   \ddot{a}_0 > 0.
\eeq
Here and in what follows  we denote $\dot{f} = df/d \tau$.

Indeed, using the relation $d\tau/d t = \hat{H}^A$ (see
(\ref{2.1})), we get
 \beq{2.6a}
    \dot{a_0} = \frac{d t}{d \tau} \frac{d a_0}{d t} =
        \frac{2|A||P| t}{\hat{H}},
  \eeq
and \beq{2.6b}
    \ddot{a_0} = \frac{d t}{d \tau} \frac{d}{d t}
    \frac{da_0}{d \tau} = \frac{2|A||P}{\hat{H}^{2 + A}} (1 + |P| t^2),
\eeq
 which certainly implies the inequalities in (\ref{2.6}).

Let us consider a variation of the effective constant $G$. In
 Jordan's frame the 4-dimensional gravitational ``constant'' is
 \beq{2.7}
      G = {\rm const}  \prod\nolimits_{i=1}^{4}
            ( a_{i}^{-d_i}) = \hat{H}^{2A} t^{-1},
 \eeq
where
 \beq{2.7a }
        a_1 = \hat{H}^{A - B_1 - B_2} t, \quad
        a_2 = \hat{H}^{A  - B_1 - B_2}, \quad
        a_3 = \hat{H}^{A - B_1}, \quad  a_4 = \hat{H}^{A - B_2}
 \eeq
are the scale factors of the ``internal'' spaces $M_1, \dots,
M_4$, respectively.

The dimensionless variation of $G$ reads
 \beq{2.11}
       \delta = \dot{G}/(GH) = 2 + \frac{1-|P| t^2}{2 A|P| t^2},
 \eeq
where
 \beq{2.12}
       H = \frac{\dot{a}_0}{a_0}
 \eeq
 is the Hubble parameter of our space. It follows from
(\ref{2.11}) that the function $G({\tau})$ has a minimum at the
point $\tau_0$ corresponding to $t_0$, where
 \beq{2.9}
      t_{0}^2 = \frac{|P|^{-1}}{1 +4 |A|}.
 \eeq
At this point, $\dot G$ is zero.

The function $G({\tau})$ monotonically decreases from $+ \infty$ to $G_0
 = G(\tau_0)$ for $\tau \in (0, \tau_0)$ and  monotonically increases from
 $G_0$ to $+ \infty$ for $\tau \in (\tau_0, \bar{\tau}_1)$. Here
 $\bar{\tau}_1 = +\infty $ for the case (i) and  $\bar{\tau}_1 = \tau_1$ for
the case (ii).

We  consider only solutions with accelerated expansion of our
space and small enough variations of the gravitational constant
obeying the  experimental constraint \cite{Hel,Dic}

 \beq{2.10}
       |\delta| < 0.1.
 \eeq
Here, as in the model with two curvatures \cite{DIKM}, $\tau$ is restricted
to a certain range containing $\tau_0$. It follows from (\ref{2.11}) that in
the asymptotical regions (\ref{2.4i}) and (\ref{2.4ii}) $\delta \to 2$,
which is unacceptable due to the experimental bounds (\ref{2.10}).  This
restriction is satisfied for a range containing the point $\tau_0$ where
$\delta = 0$.

 Calculating  $\dot G$, in the linear approximation near $\tau_0$, we get
the following approximate relation for the dimensionless parameter
of relative $G$ variation:
 \beq{2.13}
         \delta  \approx  (8 + 2 |A|^{-1}) H_0 (\tau - \tau_0),
 \eeq
where $H_0 = H(\tau_0)$ (compare with an analogous relation in
 \cite{DIKM}). This relation  gives approximate bounds for values
of the time variable $\tau$ allowed by the restriction on $\dot
 G$. A derivation of this is given in Appendix.

The solutions under consideration with $P < 0$, $d_1 = 1$ and $d_0
 = 3$ take place when the configuration of branes, the matrix
 $(h_{\alpha\beta})$ and the dilatonic coupling vectors
 $\lambda_a$, obey the relations (\ref{1.7}) and (\ref{1.10}) with
 $K_s < 0$. This is  possible when $(h_{\alpha\beta})$ is not
positive-definite, otherwise  all $K_s > 0$. Thus, there should be
at least one scalar field with negative kinetic term (i.e.,
 a phantom scalar field).

 \section{Example: a model with two phantom fields}

Let us consider the following example:   $l = 2$,
 $(h_{\alpha\beta})=-(\delta_{\alpha\beta})$, i.e. there are two
 phantom scalar fields. Due to (\ref{1.3}),
  $d(I_1) = N_1 - 1 = 1 + d_2 + d_3$ and
  $d(I_2) = N_2 - 1 = 1 + d_2 + d_4$.

Then the relations (\ref{1.7}) and (\ref{1.10}) read
   \bear{3.1}
    \vec{\lambda}_a^2 =
     (N_a - 1) \bigl( 1+ \frac{N_a - 1}{2-D}\bigr) - K_a > 0,
   \ear
  $a =1,2$, and
   \beq{3.2}
    \vec{\lambda}_1 \vec{\lambda}_2 =
     1 + d_2 - \frac{(N_1 -1)(N_2 -1)}{D - 2} +\frac{1}{2} K_2,
 \eeq
where $K_1 < 0$ and $K_2 = K_1 k$, $k =1,2,3$ for  Lie algebras
 $A_{2}$, $C_{2}$ and $G_2$, respectively. Here we have used the
relation for brane intersection: $d(I_1 \cap I_2) = 1 + d_2$.

The relations (\ref{3.1}) and (\ref{3.2}) for  are compatible
 for small enough $K_1 \in (-\infty, K_0)$, $K_0 \leq 0$,
 since it may be verified that they imply (for $K_1 \leq K_0$)

 \beq{3.3}
    \frac{\vec{\lambda}_1 \vec{\lambda}_2}
        {|\vec{\lambda}_1| |\vec{\lambda}_2|} \in (-1, +1)
 \eeq
 i.e., the vectors $\vec{\lambda}_1$ and $\vec{\lambda}_2$,
belonging to the Euclidean space $\R^2$, and obeying the relations
 (\ref{3.1}) and (\ref{3.2}), do exist. The left-hand side of
 (\ref{3.3}) gives $\cos \theta$, where $\theta$ is the angle
between these two vectors. For the special case $k=1$, $N_1 =N_2$,
 $d_3 =d_4$ considered in  \cite{AIM-07}  $K_0 = 0$.

 Now we compare the $A$ parameters corresponding to  different  Lie algebras
 $A_{2}$, $C_{2}$ and $G_2$, when the parameter $K_1$ and factor space dimensions
 $d_2, d_3, d_4$ are fixed.  We get from the definition (\ref{1.14A})

  \beq{3.4}
       A = A_{(k)}= \frac{1}{K_1 (D-2)} (n_1 d(I_1) + k^{-1} n_2 d(I_2)),
  \eeq
 or, more explicitly, (see (\ref{1.7n}))
  \bear{3.5a}
      A_{(1)} = \frac{1}{K_1 (D-2)}  (2 d(I_1) + 2 d(I_2)),
    \\  \label{3.5b}
       A_{(2)} = \frac{1}{K_1 (D-2)} (3 d(I_1) + 2 d(I_2)),
    \\  \label{3.5c}
       A_{(3)}= \frac{1}{K_1 (D-2)}  (6 d(I_1) + \frac{10}{3}
       d(I_2)),
  \ear
 for  Lie algebras  $A_{2}$, $C_{2}$ and $G_2$, respectively.
 Here $K_1 < K_0 \leq 0$.

    Hence,
  \beq{3.6}
        |A_{(1)}| < |A_{(2)}| < |A_{(3)}|.
   \eeq

   Due to relation (\ref{2.13}) for dimensionless parameter of
   relative variation of $G$ calculating in the leading approximation
   when $(\tau - \tau_0)$ is small, we get for approximate values
   of $\delta$: $\delta_{(1)}^{ap} > \delta_{(2)}^{ap} >
   \delta_{(3)}^{ap}$ that means that the variation of $G$
   (calculated near $\tau_0$)   decreases  in a sequence
   of Lie algebras $A_{2}$, $C_{2}$ and $G_2$, but
   the allowed interval $\btu \tau = \tau - \tau_0$ (
    obeying $|\delta| < 0.1$)  increases in a sequence
    of Lie algebras $A_{2}$, $C_{2}$ and $G_2$.
   This effect could be strengthen (even drastically) when $|K_1|$ becomes
   larger.   We note that for $|K_1| \to + \infty$ we get a strong
   coupling limit   $\vec{\lambda}_a^2 \to + \infty$, $a = 1,2$.

 \section{Conclusions}

We have considered  $S$-brane solutions with two non-composite
intersecting electric branes and a set of $l$ scalar fields. The
solutions contain five factor spaces, and the first one, $M_0$, is
interpreted as our 3-dimensional space.  The intersection rules
for branes correspond to the Lie algebras $A_2$, $C_2$ and $G_2$.

Here,  as in the  model with two nonzero curvatures
 \cite{DIKM}, we have found that there exists a time interval where
accelerated expansion of our 3-dimensional space is compatible
with a small enough value of $\dot{G}/G$ obeying the experimental
bounds. This interval contains  a point of minimum of the function
$G(\tau)$ denoted as $\tau_0$.

  We have analyzed special  solutions with two phantom scalar fields.
  We  have shown that in the vicinity of the point  $\tau_0$
 the time variation of $G(\tau)$ (calculated in the linear approximation)
 decreases  in the sequence of Lie algebras $A_{2}$, $C_{2}$ and $G_2$.
 Thus, this treatment justifies the consideration of different
 non-simply laced Lie algebras (such as $C_{2}$ and $G_2$) that
 usually do not appear for stringy-inspired solutions \cite{IMJ}.

 \renewcommand{\theequation}{\Alph{section}.\arabic{equation}}
 \renewcommand{\thesection}{}
 \setcounter{section}{0}

 \section{Appendix}

Here we give a derivation of the relation (\ref{2.13}) for an
approximate value of the dimensionless parameter of relative
variation of $G$.

We start with the relation (\ref{2.11}) written in the following
form
  \beq{A.1}
       \delta = \dot{G}/(GH) = \frac{t^2- t^2_0}{|A P| t_0^3},
   \eeq
where  $H = \dot{a}_0/a_0$ is the Hubble parameter and
 $t_{0}$ is defined in  (\ref{2.9}). (We recall
 that here and in what follows  $\dot{f} = df/d \tau$.)
 In the vicinity of the point
 $t_0$ we get in linear approximation
  \beq{A.2}
       \delta \approx  \frac{\btu t}{|A P| t_0^3}.
   \eeq
Using the synchronous time variable  $\tau = \tau(t)$ we get
    \beq{A.2a}
   \delta \approx  \left(\frac{dt}{d \tau}\right)_0 \frac{\btu \tau}{|A P| t_0^3}=
    \frac{\btu \tau}{\hat{H}^A_0 |A P| t_0^3}
   \eeq
 ($d\tau/d t = \hat{H}^A$). Here the subscript "0" refers to
 $t_0$. For the Hubble parameter we get  from (\ref{2.3})
 and (\ref{2.6a})
  \beq{A.3}
    H_0 = \left(\frac{\dot{a}_0}{a_0}\right)_0 =
          \frac{2|A||P| t_0}{\hat{H}^{A+1}_0}.
  \eeq
  Then, it follows from (\ref{A.2a}) and   (\ref{A.3}) that
  \beq{A.4}
   \delta \approx   \frac{\hat{H}_0 }{2 A^2 P^2 t_0^4} H_0 \btu
   \tau.
  \eeq
  Since (see (\ref{1.5}) and (\ref{2.9}))
 \beq{A.5}
  \hat{H}_0 = 1 - \frac{1}{1 +4 |A|} = \frac{4 |A|}{1 +4 |A|}
 \eeq
 the pre-factor in (\ref{A.4}) reads (see (\ref{2.9})):
  \beq{A.6}
   \Pi = \frac{\hat{H}_0 }{2 A^2 P^2 t_0^4} = 8 + 2 |A|^{-1}
  \eeq
 and  we are led to the relation
 \beq{A.7}
         \delta  \approx  \Pi  H_0 (\tau - \tau_0),
 \eeq
 coinciding with (\ref{2.13}).

 \begin{center}
 {\bf Acknowledgments}
 \end{center}

 This work was supported in part by the Russian Foundation for
 Basic Research grant  Nr. $07-02-13624-ofi_{ts}$.

\end{document}